\documentclass[10pt,a4paper]{article}

\usepackage[margin=1.0in]{geometry}
\usepackage{authblk}
\usepackage{amsmath}
\usepackage{graphicx}
\usepackage{parskip}
\usepackage{hyperref}
\newcommand\dd{\mathrm{d}}
\newcommand\ee{\mathrm{e}}

\title{Compact stars in $f(T)$ extended theory of gravity}
\author{Sa{\v s}a Iliji{\'c}\thanks{E-mail: {\texttt sasa.ilijic@fer.hr}} }
\author{Marko Sossich\thanks{E-mail: {\texttt marko.sossich@fer.hr}} }
\affil{University of Zagreb, Faculty of Electrical Engineering and Computing,\\
       Department of Applied Physics, Unska 3, HR-10\,000 Zagreb, Croatia}
\date{Sep 20, 2018}

\begin{document}

\maketitle

\begin{abstract}

We consider static spherically symmetric
self-gravitating configurations of the perfect fluid
within the framework of the torsion-based extended theory of gravity.
In particular, we use the covariant formulation of $f(T)$ gravity
with $f(T) = T + \frac{\alpha}{2} T^2$,
and for the fluid we assume the polytropic equation of state
with the adiabatic exponent $\Gamma = 2$.
The constructed solutions have a sharply defined radius
[as in General Relativity (GR)]
and can be considered as models of nonrotating compact stars.
The particle number--to--stellar radius curves
reveal that with positive (negative) values of $\alpha$
smaller (greater) number of particles
can be supported against gravity then in GR.
For the interpretation of the energy density and the pressure within the star
we adopt the GR picture where the effects due to nonlinearity of $f(T)$
are seen as a $f(T)$ fluid,
which together with the polytropic fluid
contributes to the effective energy momentum.
We find that sufficiently large positive $\alpha$
gives rise to an abrupt sign change (phase transition)
in the energy density and in the principal pressures of the $f(T)$ fluid,
taking place within the interior of the star.
The corresponding radial profile of the effective energy density
is approximately constant over the central region of the star,
mimicking an incompressible core.
This interesting phenomenon is not found
in configurations with negative $\alpha$.

\end{abstract}


\section{Introduction}

Since the early days of General Relativity (GR), it has been known
that replacing the Ricci scalar $R$ in the Einstein-Hilbert action
with the torsion scalar $T$ yields a theory of gravity
of which the equations of motion are equivalent to those of GR.
The torsion-based variant of the theory is known
as the teleparallel equivalent of General Relativity
\cite{Aldrovandi:2013wha}. 
However, if the gravitational action is extended
by allowing for terms of the form $f(R)$ or $f(T)$,
where $f$ is a nonlinear function,
the resulting curvature-based and torsion-based
extended theories of gravity become grossly different.
The curvature-based $f(R)$ gravity
has been thoroughly studied over the past decades
\cite{Sotiriou:2008rp}.
Its most remarkable feature
is the occurrence of fourth order derivatives of the metric
in the resulting equations of motion.
In contrast to this,
the equations of motion of the torsion-based $f(T)$ theory
involve only the usual second order derivatives of the tetrad fields.
The features of $f(T)$ theory have been
widely studied in the cosmological setting,
see e.g., Refs.~\cite{%
   Cai:2015emx,
   Yang:2010hw,
   Myrzakulov:2010vz,
   Awad:2017yod,
   Awad:2017ign,
   Bahamonde:2017ize,
   Nojiri:2017ncd,Nunes:2018xbm}, 
and some constraints on the theory
have been determined through considerations
of the motion of planets in the Solar System,
see e.g., Refs.~\cite{%
   Farrugia:2016xcw,
   Iorio:2015rla,
   Qi:2017xzl},
while investigations in the arena of static spherical symmetry,
in particular those considering stellar structure,
are somewhat fewer in number \cite{%
   Boehmer:2011gw,
   Daouda:2011rt,
   Kpadonou:2015eza,
   Abbas:2015yma,
   2015Ap&SS.359...57A,
   Momeni:2016oai,
   DeBenedictis:2018wkp,
   2017PDU....16...34B} 
The structure of the theory itself has also been investigated
\cite{deAndrade:2000kr,Ferraro:2018tpu}.
An important problem that was found
in the early formulations of the $f(T)$ theory
is the lack of the Lorentz invariance
in the sense that the equations of motion are in general not invariant
with respect to the choice of the metric compatible tetrad;
see e.g., Refs.~\cite{Li:2010cg,Sotiriou:2010mv}.
In recent years, a lot of effort has been invested into
understanding and solving this problem.
Some of the promising results
are in the form of the Hamiltonian formalism \cite{Li:2011rn},
the null tetrad approach \cite{Bejarano:2014bca},
the Lagrange multiplier formulation \cite{Blagojevic:2000pi,Nester:2017wau},
and the covariant formulation \cite{%
   Krssak:2015oua,
   Golovnev:2017dox,
   Hohmann:2018rwf}.
In this work, we will rely on the covariant formulation of $f(T)$ gravity
in the form proposed by Kr{\v s}{\v s}{\'a}k
and Saridakis in Ref.~\cite{Krssak:2015oua}.

In GR, the models of nonrotating stars
can be constructed within static spherical symmetry
as solutions to the well-known Tolman-Oppenheimer-Volkov equations.
One must also assume a matter model,
which is often a perfect fluid subject to some equation of state.
In general,
the energy density and the pressure of the fluid may take up all space,
while if they strictly vanish
outside of a spherical surface of some finite radius,
the models are said to represent compact stars surrounded by vacuum.
Within GR, the phenomenology of these solutions is well understood.
The most important results include
proofs of existence of solutions with certain classes of equations of state
\cite{Rendall:1990hg,BaumRend93},
upper bounds on the stellar mass--to--surface radius ratio
that is also a measure of the ``compactness'' of a compact object
\cite{Buch59,Andreasson:2007ck},
as well as the relation among the stellar mass--to--surface radius curves
and the dynamical stability of the models \cite{HTWW65}.
Our main goal in this paper is to obtain numerical evidence
for the existence of similar solutions
within the torsion-based extended theory of gravity.
We intend to construct an analog
of the stellar mass--to--radius curves of GR
and search for possible signatures of the extended theory in the solutions.
We will use the simplest possible variant of the $f(T)$ theory
where the nonlinear function $f$ involves a term which is quadratic in $T$
[this variant of $f(T)$ can be seen as an analog
of the Starobinsky $f(R)$ model in curvature-based theory
\cite{Starobinsky:1980te},
which has also become known as the $R^2$-gravity theory].
For the equation of state of the fluid, we adopt
the polytropic equation of state with the adiabatic exponent $\Gamma = 2$,
since it represents the most stiff model of matter
which satisfies the most important physical constraints
in all pressure regimes.

The paper is organized as follows.
In Sec.~\ref{sec:foft} we briefly review the equations of motion
of the covariant $f(T)$ theory in static spherical symmetry.
In Sec.~\ref{sec:eos} we introduce the polytropic equation of state,
and in Sec.~\ref{sec:num} we discuss the rescaling of the equations,
the boundary conditions that we impose, and the numerical procedure.
In Secs.~\ref{sec:nr} and~\ref{sec:profiles},
we discuss the particle number--to--stellar radius curves
and the radial profiles of the energy density and the pressures.
We conclude in Sec.~\ref{sec:conclusion}.
Geometrized units $c=1=G$ are used throughout the paper.


\section{Static spherical symmetry in $f(T)$ \label{sec:foft}}

The $f(T)$-gravity theory follows from the action written as
\begin{equation} \label{eq:action}
S = \int \left(
       \frac{f(T)}{16\pi} + \mathcal{L}_{\mathrm{matter}}
    \right) \det[h^a{}_\mu] \; \dd^{4}x ,
\end{equation}
where $f$ is in general a nonlinear function of the torsion scalar $T$,
$\mathcal{L}_{\mathrm{matter}}$ is the Lagrangian density due to matter fields,
and $h^a{}_\mu$ is the tetrad,
which is the dynamical degree of freedom of this theory.
We use latin symbols for the tetrad indices
and greek symbols for the spacetime indices.
The tetrad satisfies the metric compatibility condition
$h^a{}_\mu h^b{}_\nu g^{\mu\nu} = \eta^{ab}$,
where $g^{\mu\nu}$ is the spacetime metric tensor
and $\eta^{ab}$ is the Minkowski tensor.
If $f(T)=T$, the resulting equations of motion
are equivalent to those of GR, while if $f$ is nonlinear in $T$,
the equation of motion can be written as
\cite{Aldrovandi:2013wha,Krssak:2015oua}
\begin{equation} \label{eq:genericeom}
h^{-1} h^{a}{}_{\mu} \partial_{\sigma}
     \left( h \frac{\dd f(T)}{\dd T} S_{a}{}^{\nu\sigma} \right)
- \frac{\dd f(T)}{\dd T} T_{\alpha\beta\mu} S^{\alpha\beta\nu}
+ \frac12 f(T) \delta_{\mu}{}^{\nu}
+ \frac{\dd f(T)}{\dd T}
          S_{a}{}^{\alpha\nu}
          h^{b}{}_{\mu}
          \omega^{a}{}_{b\alpha}
 = 8\pi \mathcal{T}_{\mu}{}^{\nu}.
\end{equation}
In the above equation,
$\mathcal{T}_{\mu}{}^{\nu}$ is the usual energy-momentum tensor, and
\begin{equation} \label{eq:torsiontensor}
T^{\alpha}{}_{\beta\gamma} =
h_{a}{}^{\alpha} \left( \partial_{\beta}  h^{a}{}_{\gamma}
                      - \partial_{\gamma} h^{a}{}_{\beta} \right)
+ h_{a}{}^{\alpha} \omega^{a}{}_{b\beta}  h^{b}{}_{\gamma}
- h_{a}{}^{\alpha} \omega^{a}{}_{b\gamma} h^{b}{}_{\beta}
\end{equation}
is the torsion tensor.
The quantity $\omega^{a}{}_{b\alpha}$
is the inertial spin connection which is,
in the covariant formulation of $f(T)$ gravity \cite{Krssak:2015oua},
determined from the requirement
that the torsion tensor vanishes in the flat-space limit of the metric.
The tensors
\begin{equation} \label{eq:ktensor}
    K_{\alpha\beta\gamma} =
     \frac12 \left(  T_{\alpha\gamma\beta}
     +  T_{\beta\alpha\gamma} +  T_{\gamma\alpha\beta} \right)
\end{equation}
and
\begin{equation} \label{eq:stensor}
    S_{\alpha\beta\gamma} =  K_{\beta\gamma\alpha}
      + g_{\alpha\beta} \,  T_{\sigma\gamma}{}^{\sigma}
      - g_{\alpha\gamma} \,  T_{\sigma\beta}{}^{\sigma}
\end{equation}
are known as the contorsion tensor and the modified torsion tensor.
The torsion scalar,
\begin{equation}
T =T^{\alpha\beta\gamma} S_{\alpha\beta\gamma},
\end{equation}
is then defined as the contraction of the torsion tensor
with the modified torsion tensor.

The specific form of $f(T)$ which we will be using is
\begin{equation}
f(T) = T + \frac{\alpha}2 T^2 .
\end{equation}
With the above choice of $f(T)$,
the left-hand side of the equation of motion (\ref{eq:genericeom})
can be written as the sum of the Einstein tensor $G_{\mu}{}^{\nu}$
built from the Christoffel connection
and the additional term proportional to $\alpha$
which we denote with the tilded symbol $\tilde G_{\mu}{}^{\nu}$.
We then find it convenient to adopt the ``GR picture'' of the equation of motion
by transferring $\tilde G_{\mu}{}^{\nu}$ to the right-hand side
and writing the equation as
\begin{equation} \label{eq:tildedeom}
G_{\mu}{}^{\nu} = 8\pi \left( {\mathcal{T}}_{\mu}{}^{\nu}
                + \tilde {\mathcal{T}}_{\mu}{}^{\nu} \right),
\end{equation}
where the quantity
\begin{equation} \label{eq:tildet}
\tilde{\mathcal{T}}_{\mu}{}^{\nu} = - \frac1{8\pi} \tilde G_{\mu}{}^{\nu}
\end{equation}
can be interpreted as the contribution to the standard energy-momentum tensor
arising from the nonlinear term in $f(T)$.
We will refer to (\ref{eq:tildet})
as the energy-momentum tensor of the ``$f(T)$ fluid''.

As we are intending to construct the models of static nonrotating stars,
we now assume static spherical symmetry.
We use spherical coordinates $x^{\mu} = (t,r,\vartheta,\varphi)$
and write the metric tensor as
\begin{equation} \label{eq:metric}
g_{\mu\nu} = \mathrm{diag}\left(
\mathrm{e}^{2\Phi(r)}, -\mathrm{e}^{2\Lambda(r)},-r^2,-r^2 \sin^2\vartheta
\right).
\end{equation}
The tetrad compatible with the above metric can be chosen as
\begin{equation} \label{eq:tetrad}
h^a{}_{\mu} = \mathrm{diag}\left(
\ee^{\Phi(r)}, \ee^{\Lambda(r)},r,r \sin\vartheta
\right).
\end{equation}
The condition that the torsion tensor (\ref{eq:torsiontensor})
vanishes in the flat spacetime limit,
which is obtained by letting $\Phi\to 0$ and $\Lambda\to 0$,
gives the inertial spin connection of which the nonzero components are
$\omega^{\hat r}{}_{\hat \vartheta \vartheta}
= - \omega^{\hat \vartheta}{}_{\hat r \vartheta} = -1$,
$\omega^{\hat r}{}_{\hat \varphi \varphi}
= - \omega^{\hat \varphi}{}_{\hat r \varphi} = - \sin\vartheta$, and
$\omega^{\hat \vartheta}{}_{\hat \varphi \varphi}
= - \omega^{\hat \varphi}{}_{\hat \vartheta \varphi} = -\cos\vartheta$
(here, the coordinate labels are used as indices,
and the tetrad indices are distinguished from the spacetime indices
with the hat symbol).
After lengthy but straightforward manipulations,
one obtains the torsion scalar
\begin{equation}
T = {2}{r^{-2}} \ee^{-2\Lambda}
    \left( \ee^\Lambda - 1 \right)
    \left( \ee^\Lambda - 2 r \Phi' - 1 \right),
\end{equation}
where the prime denotes differentiation with respect to $r$.
The nonzero components of the Einstein tensor are
\begin{alignat}{1}
G_t{}^t & = r^{-2} \left( 1 - \ee^{-2\Lambda}
            \left( 1 - 2 r \Lambda' \right) \right) , \\
G_r{}^r & = r^{-2} \left( 1 - \ee^{-2\Lambda}
            \left( 1 + 2 r \Phi' \right) \right) , \\
G_\vartheta{}^\vartheta & = G_\varphi{}^\varphi
            = r^{-1} \ee^{-2\Lambda} \left(
            \left( \Lambda' - \Phi' \right)
            \left( 1 + r \Phi' \right) - r \Phi'' \right) ,
\end{alignat}
while the nonzero terms on the left-hand side
of the equation of motion (\ref{eq:genericeom})
that are proportional to $\alpha$
are given by somewhat more complicated expressions:
\begin{alignat}{1}
\tilde G_t{}^t & = \alpha r^{-4} \ee^{-4\Lambda}
   \left( \ee^\Lambda - 1 \right) \big(
      4 r \Lambda' \left(
         3 \left( \ee^\Lambda - 1 \right) +
         2 \left( \ee^\Lambda - 3 \right) r \Phi' \right) +
\notag \\ & \qquad
      \left( \ee^\Lambda - 1 \right)
      \left( \left( \ee^\Lambda -1 \right) \left( \ee^\Lambda -5 \right) -
      4 r^2 \left( \Phi'^2 + 2 \Phi'' \right) \right) \big) , \\
\tilde G_r{}^r & = \alpha r^{-4} \ee^{-4\Lambda}
                   \left( \ee^\Lambda - 1 \right)
                   \left( \ee^\Lambda - 1 - 2 r \Phi' \right)
     \left( \left( \ee^\Lambda - 1 \right) \left( \ee^\Lambda + 3 \right)
            + 2 \left( \ee^\Lambda - 3 \right) r \Phi' \right) , \\
\tilde G_\vartheta{}^\vartheta & = \tilde G_\varphi{}^\varphi =
   \alpha r^{-4} \ee^{-4\Lambda} \big(
\notag \\ & \qquad 
2 r \Lambda' \big(
   3 \left( \ee^\Lambda - 1 \right)^2 +
   3 \left( \ee^\Lambda - 1 \right) \left( \ee^\Lambda - 3 \right) r \Phi' -
   2 \left( 2 \ee^\Lambda - 3 \right) r^2 \Phi'^2 \big) -
\notag \\ & \qquad
   \left( \ee^\Lambda - 1 \right) \big(
   \left( \ee^\Lambda - 1 \right)^2 \left( \ee^\Lambda + 3 \right) +
   2 \big( 3 \left( \ee^\Lambda - 1 \right) r^2 \Phi'^2 -
   2 r^3 \Phi'^3 +
\notag \\ & \qquad
   3 \left( \ee^\Lambda - 1 \right) r^2 \Phi'' +
   \left( 1 + \ee^\Lambda - 2 \ee^{2\Lambda} - 4r^2 \Phi'' \right) r \Phi'
   \big) \big) \big) .
\end{alignat}
The standard energy-momentum tensor of the perfect fluid
in the static spherical symmetry has the well-known diagonal form
\begin{equation} \label{eq:stdEnergyMomentum}
\mathcal T_{\mu}{}^{\nu}
= \mathrm{diag}(\rho,-p,-p,-p),
\end{equation}
where $\rho$ is the energy density
and $p$ is the isotropic pressure of the fluid.
The complete equations of stellar structure can now be written as
\begin{equation} \label{eq:eom3}
G_t{}^t = 8\pi ( \rho + \tilde \rho ), \qquad
G_r{}^r = - 8\pi ( p + \tilde p ), \qquad
G_\vartheta{}^\vartheta = - 8\pi ( p + \tilde q ),
\end{equation}
where the quantities
\begin{equation} \label{eq:threeMonkeys}
\tilde \rho = - \frac{1}{8\pi} \tilde G_t{}^t, \qquad
\tilde p = \frac{1}{8\pi} \tilde G_r{}^r, \qquad
\tilde q = \frac{1}{8\pi} \tilde G_\vartheta{}^\vartheta,
\end{equation}
are the energy density, the radial, and the transverse pressure
of the $f(T)$ fluid.
It is also important to note that,
while the quantities (\ref{eq:threeMonkeys}) vanish in the Minkowski spacetime,
they are nonzero in the general Schwarzschild spacetime;
i.e., the Schwarzschild metric
is not a vacuum solution of the nonlinear $f(T)$ theory of gravity
\cite{DeBenedictis:2016aze}.
This means that at the surface of the star,
where the components of the fluid energy-momentum tensor
(\ref{eq:stdEnergyMomentum}) vanish
and where the interior metric joins the exterior vacuum metric,
the quantities (\ref{eq:threeMonkeys}) may have nonzero values.


\section{Equation of state of the polytropic fluid \label{sec:eos}}

The equation of state (EoS) of a perfect fluid is in general specified
as the dependence of the fluid pressure $p$ on the particle number density $n$
and the entropy per particle $s$, i.e., by specifying the function $p=p(n,s)$.
Assuming the isentropic flow of the fluid,
the entropy per particle $s$ is constant,
allowing one to write the EoS simply as $p=p(n)$.
The so-called polytropic EoS follows
form the assumption that the adiabatic exponent
$\Gamma = \left( {\partial \ln p}/{\partial \ln n} \right)_s$
is independent of the pressure.
For the EoS of the form $p=p(n)$, this implies 
$\Gamma = ({n}/{p}) ({\dd p}/{\dd n})$,
which upon integration gives
   \begin{equation} \label{eq:pn}
   p = k n^{\Gamma},
   \end{equation}
where $k$ is the integration constant.
The energy density of the polytropic fluid
can be obtained by invoking the first law of thermodynamics.
For a fluid element of proper volume $V$, it can be expressed as
$T \mathrm{d}(n s V) = \mathrm{d} (\rho V) + p \, \mathrm{d}V$;
since the number of particles in $V$ is constant,
we use $V\propto 1/n$; and since in the isentropic flow $s$ is constant,
it implies
   \begin{equation}
   \frac{\mathrm{d} \rho}{\mathrm{d} n} = \frac{\rho+p}{n}
   = \frac{\rho + k n^{\Gamma}}{n} .
   \end{equation}
Upon integration, one obtains
   \begin{equation} \label{eq:rhon}
   \rho = m n + \frac{k n^{\Gamma}}{\Gamma-1},
   \end{equation}
where the integration constant $m$ can be recognized
as the mass of a single particle.
Using (\ref{eq:pn}) and denoting $\kappa = m k^{-1/\Gamma}$,
the above relation can also be written in the form
   \begin{equation} \label{eq:releos}
   \rho = \kappa p^{1/\Gamma} + \frac{p}{\Gamma-1}
   \end{equation}
which is known as the polytropic EoS.
The square of the speed of the sound waves,
sometimes also referred to as the stiffness of the fluid,
for the fluid obeying the EoS (\ref{eq:releos}) is given by
   \begin{equation}
   c_{\mathrm{sound}}^2 = \left( \frac{\partial p}{\partial \rho} \right)_s = 
   \frac{ \Gamma-1 }{
    1 + ( 1 - 1/\Gamma ) \kappa p^{-( 1 - 1/\Gamma )} } .
   \end{equation}
We see that $c_{\mathrm{sound}}^2$ is monotonically increasing with pressure
and is bounded from above by $\Gamma - 1$.
This implies that the polytropic fluids with $\Gamma \le 2$ are causal
(sound propagates at subluminal speeds) at all pressures.
It is also easy to show that the pressure--to--energy density ratio
   \begin{equation}
   \sigma = \frac{p}{\rho} = \frac{ \Gamma - 1 }{
      1 + ( \Gamma - 1 ) \kappa p^{-( 1 - 1/\Gamma )} }
   \end{equation}
is bounded from above with $\Gamma-1$.
This means that the dominant energy condition (DEC),
which requires that $p<\rho$,
is satisfied at all pressures if $\Gamma\le2$.
The limiting case of the polytropic fluid with $\Gamma=2$
can be singled out as the maximally stiff model of matter
capable of withstanding arbitrarily large pressures
without violating the basic principles (causality and DEC).
We will therefore use $\Gamma = 2$ in our models of compact objects.


\section{Numerical procedure \label{sec:num}}

In principle, the system of three coupled
ordinary differential equations (\ref{eq:eom3}),
the EoS of the polytropic fluid (\ref{eq:releos}),
and the set of appropriate boundary conditions (to be discussed below)
determine the structure of the polytropic star or polytrope
in $f(T) = T + \frac{\alpha}{2} T^2$ gravity.
The parameter space is four dimensional
and the parameters can be chosen as
$\kappa>0$ and $\Gamma>1$ appearing in the EoS,
the value of the fluid pressure--to--energy density ratio
at the center of the star $\sigma_0 = p(0)/\rho(0)$,
and the coefficient $\alpha$ of the quadratic term in $f(T)$.
The variables to be solved for can be chosen
as the metric profile functions $\Phi(r)$ and $\Lambda(r)$
and the fluid pressure profile function $p(r)$.
However, some elementary transformations make the system
more adequate for the numerical treatment.
The parameters $\kappa$ and $\gamma$
allow one to define the parameter
   \begin{equation}
   \lambda = \kappa^{-\Gamma/2(\Gamma-1)}
   \end{equation}
which has the dimension of length
and which can be used as the length scale for the problem.
In this way, the dimension of the parameter space reduces by one,
and we chose to work with three dimensionless parameters,
   \begin{equation}
   a = \lambda^{-2} \alpha, \qquad
   \Gamma, \qquad
   \sigma_0 = p(0)/\rho(0).
   \end{equation}
For the variables to solve for, we chose the functions
   \begin{equation}
   \Phi'(r), \qquad
   \Lambda(r), \qquad
   \sigma(r) = p(r)/\rho(r).
   \end{equation}
[Note that the function $\Phi(r)$
does not appear in the equations of stellar structure,
which means that it is sufficient to solve for $\Phi'(r)$.]
The transformation of the system of equations is straightforward,
but since the resulting expressions are rather cluttered,
they are not shown explicitly.

The boundary conditions appropriate for the problem
can be specified at the center of symmetry and at the surface of the star.
At the center of symmetry, we require
   \begin{equation} \label{eq:bc0}
   \Phi'(0) = 0, \qquad \Lambda(0) = 0, \qquad \sigma(0) = \sigma_0.
   \end{equation}
The surface of the star is defined as the hypersurface $r=R$
at which the energy density of the fluid vanishes
and the interior spacetime joins with the vacuum exterior.
The boundary condition reflecting this is $\sigma(R) = 0$.
However, as the value of $R$ is \emph{a priori} unknown,
$R$ must be treated as an eigenvalue of the problem.
Technically, this is done by introducing
the dimensionless radial coordinate $x=r/R$
which maps the problem onto the compact domain $x\in[0,1]$.
The boundary condition at the stellar surface is then
   \begin{equation} \label{eq:bc1}
   \sigma(1) = 0 .
   \end{equation}
The eigenvalue $R$ is obtained through the numerical solution
of the boundary value problem (we use the tool \textsc{colsys} \cite{COLSYS}),
and the validity of each converged solution is confirmed
by means of the \emph{a posteriori} initial value integration.
We also made sure that the results we obtain within GR (for $a=0$)
reproduce the results from the literature,
e.g., Refs.~\cite{1964ApJ...140..434T, Bludman:1972zz}.

It should be noted that within the framework of the $f(T)$ gravity
one cannot unambiguously compute the mass of the star.
While in GR the mass of a compact (finite radius) star
computed from $M=4\pi\int_0^R \rho r^2 \, \dd r$
coincides with the Arnowitt-Deser-Misner (ADM) \cite{adm62}
mass of the spacetime, in $f(T)$ gravity, this result is not known to hold.
Within the nonlinear $f(T)$-gravity theory,
as a robust measure of the amount of matter contained
within a certain stellar configuration,
one can use the total particle number $N$.
It can be obtained by integrating the particle number density
$n = \dd N / \dd V$ over the interior of the star.
Writing the proper volume element as $\dd V = 4r^2\pi \ee^{\Lambda(r)} \dd r$,
the particle number is given by
\begin{equation}
N 
= 4\pi \int_0^R n(r) \ee^{\Lambda(r)} r^2 \, \dd r ,
\end{equation}
where the particle number density $n(r)$
is to be expressed in terms of $\sigma(r)$.


\section{Particle number-to-stellar radius curves \label{sec:nr}}

\begin{figure}
\begin{center}
\includegraphics{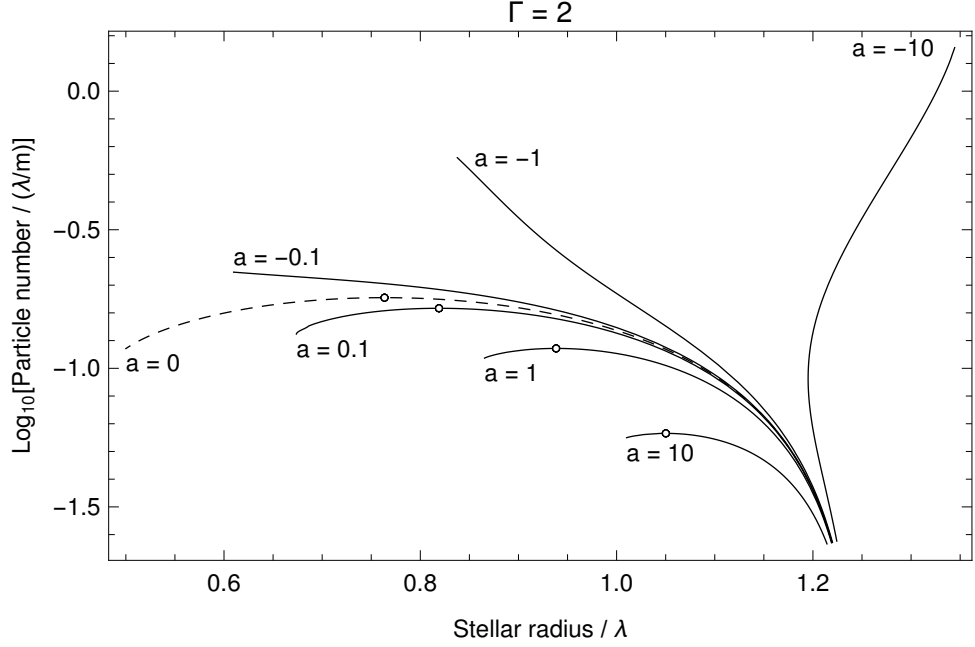}
\end{center}
\caption{\label{fig:g2_NR}
Particle number--to--stellar radius curves
for polytropes with adiabatic exponent $\Gamma = 2$
computed within GR ($a=0$, dashed line)
and within the nonlinear $f(T)$-gravity theory
($a=\pm0.1,\pm1,\pm10$, solid lines).}
\end{figure}

\begin{table}
\caption{\label{tbl:g2_NR}}
Parameters of some configurations of the $\Gamma=2$
polytropes computed within GR and within the nonlinear $f(T)$-gravity theory.
\begin{center}
\begin{tabular}{l|l|l|l|l|l}
 $a$ & $\sigma_0$ & $R/\lambda$ & $N/(\lambda/m)$ & $\mu(R)$ & Comment \\
\hline
  10   & 0.0369606 & 1.05013 & 0.0582413 & 0.117202 & Critical \\
       & 0.0404008 & 1.01023 & 0.0561978 & 0.123072 & Max.~$\sigma_0$\\
\hline
  1    & 0.0974466 & 0.938375 & 0.117965 & 0.296724 & Critical \\
       & 0.117761 & 0.865438 & 0.108835 & 0.323215 & Max.~$\sigma_0$ \\
\hline
  0.1  & 0.187386 & 0.819034 & 0.164647 & 0.519582 & Critical \\
       & 0.304203 & 0.673722 & 0.132935 & 0.604828 & Max.~$\sigma_0$ \\
\hline
0 (GR) & 0.241407 & 0.763518 & 0.179862 & 0.626006 & Critical \\
\hline
 -0.1  & 0.473319 & 0.609773 & 0.22232 & 0.906915 & Max.~$\sigma_0$ \\
\hline
 -1    & 0.451499 & 0.837695 & 0.575407 & 0.890173 & Max.~$\sigma_0$ \\
\hline
 -10   & 0.316375 & 1.34443 & 1.43548 & 0.644026 & Max.~$\sigma_0$ \\
\end{tabular}
\end{center}
\end{table}

In General Relativity,
a family of static spherically symmetric stellar models
obtained with certain EoS and corresponding to a range of values
of the central pressure--to--energy density ratio $\sigma_0$
is often represented by a curve on a mass--to--radius graph.
The typical behavior of a mass--to--radius curve is such that
as $\sigma_0$ increases starting from a sufficiently low value
the radius of the star decreases,
while the mass increases, reaches the maximum,
and decreases with a further increase of $\sigma_0$.
The configuration with maximal mass
is usually called the critical configuration
because it has been shown that all configurations with $\sigma_0$
lower than that corresponding to that configuration
are stable with respect to small radial perturbations.
If $\sigma_0$ increases beyond that critical value,
the configurations become dynamically unstable.
A similar pattern is found if instead of stellar mass $M$
one displays the particle number $N$ of the configuration.
The dashed line in Fig.~\ref{fig:g2_NR}
shows the dependence of the particle number
on the stellar radius for the GR ($a=0$) polytrope with $\Gamma = 2$.
Solid lines in Fig.~\ref{fig:g2_NR}
are the particle number--to--radius curves obtained with the same EoS,
but within the nonlinear $f(T)$ theory
with the coefficient $a=\pm 0.1, \pm 1, \pm 10$.
All curves start at $\sigma_0=0.01$ and are continued
only as far as the numerical procedure could obtain stable convergences.
The critical configurations, in cases where they exist,
are indicated by circle symbols.
Precise values of the particle number and stellar radius
for some of the solutions are given in Table~\ref{tbl:g2_NR}.
The table also contains the quantity
$\mu(R) = 1 - \ee^{-2\Lambda(R)}$,
which is computed directly from the metric
and which in GR corresponds to the ratio $2M/R$,
where $M$ is the mass of the star.
This quantity is usually taken as the measure
of the compactness of the object and is bounded from above by unity.

We first note that as $\sigma_0 \to 0$  solutions
with all values of $a$ that we computed
approach the nonrelativistic polytrope configuration
described by the solution to the Lane-Emden equation
with the polytropic index $n=1$ (see e.g., Ref.~\cite{Glen00}).
The stellar radius of this polytrope is $R=\sqrt{\pi/2}\,\lambda$.

For positive $a$, the particle number--to--radius curves that we obtained
are qualitatively similar to the curve representing the GR solutions.
In each of these curves, the critical configuration can be found.
As $a$ increases, the radius of critical configuration increases,
while the particle number decreases.
This indicates that the amount of matter
that can be supported against gravity with $a>0$ is less than in GR.

For negative $a$, if $|a|$ is sufficiently large,
the particle number--to--radius curves may develop behavior
which qualitatively differs from the behavior in GR.
Starting from the nonrelativistic regime, as $\sigma_0$ increases,
the particle number increases,
but before the critical configuration is reached,
the numerical procedure stops converging,
indicating the lack of existence of solutions beyond that value of $\sigma_0$
(see the $a=-0.1$ or $a=-1$ curve in Fig.\ \ref{fig:g2_NR}).
With still larger values of $|a|$,
the particle number--to--radius curves
reveal configurations with the minimal radius,
beyond which the radius starts to increase
(see the $a=-10$ curve in Fig.\ \ref{fig:g2_NR}).
The critical points analogous to those
that are typical for mass-to-radius curves in GR
are no longer present in particle number--to--radius curves
with sufficiently large negative parameter $a$.
We also observe that, in comparison with the GR configurations,
negative $a$ allows static configurations of much larger amounts
of matter (number of particles) described by the same polytropic EoS.
However, since we are not dealing with the full dynamic theory,
nor with linearized dynamics of the system
that would allow us to consider small perturbations,
we are at present not able to make any claims
regarding dynamical stability of any of the configurations
computed within the nonlinear $f(T)$-gravity theory.


\section{Energy density and pressure profiles \label{sec:profiles}}

\begin{figure}
\begin{center}
\includegraphics{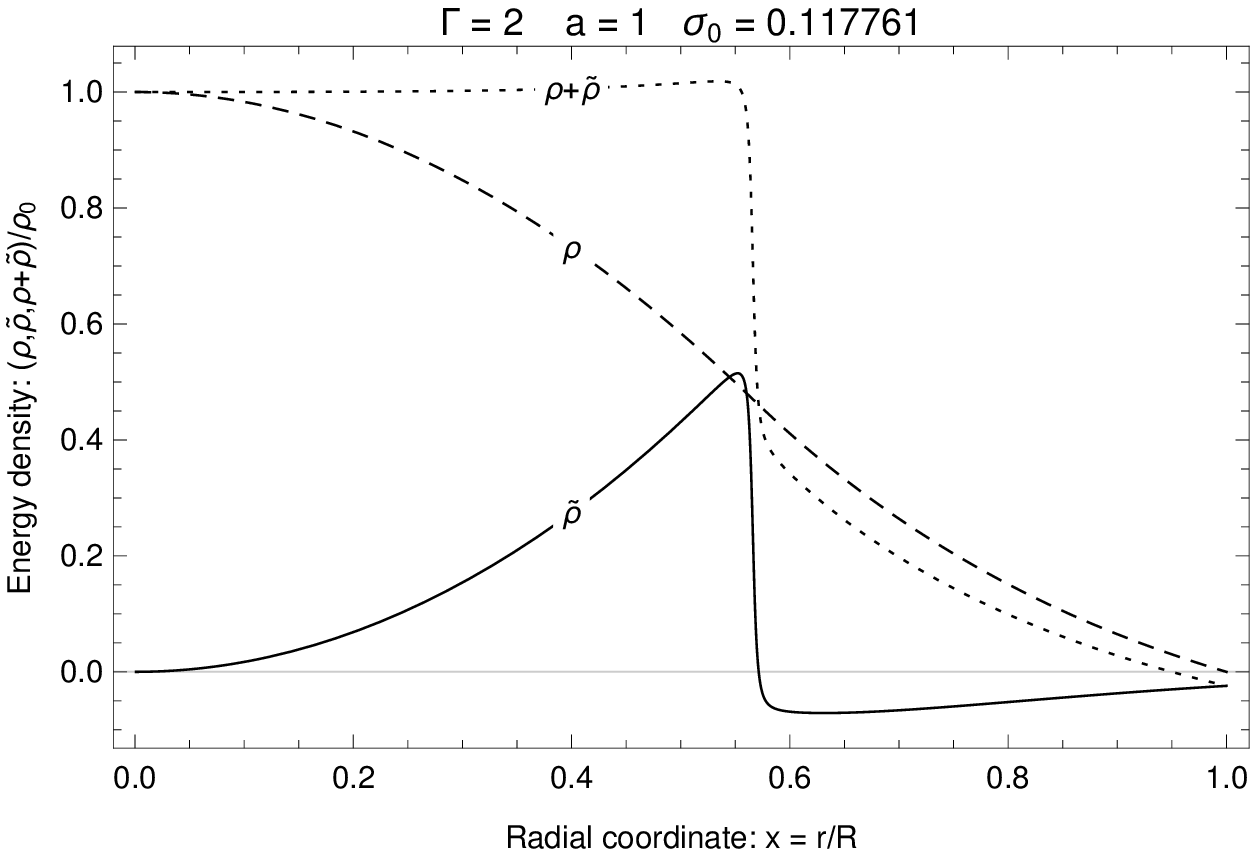}\\[1em]
\includegraphics{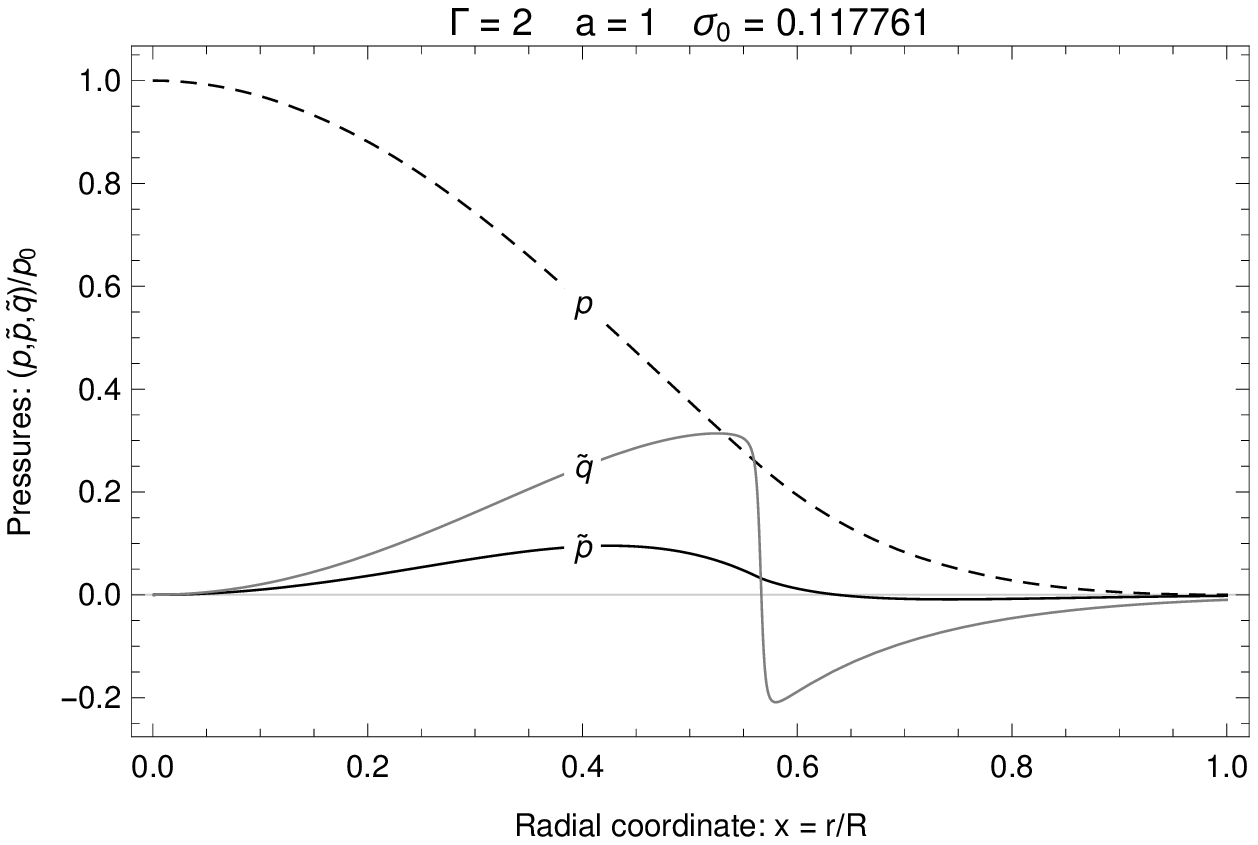}
\end{center}
\caption{\label{fig:plus}
The $\Gamma = 2$ polytrope (EoS $\rho=\kappa p^{1/2} + p$)
with maximal $\sigma_0$ in $f(T) = T + \frac{\alpha}{2} T^2$-gravity theory,
$a=\alpha\kappa^2=1$.
Upper plot: energy density $\rho$ of the polytropic fluid (dashed line),
energy density $\tilde\rho$ of the $f(T)$ fluid in the GR picture (solid line),
and effective energy density $\rho+\tilde\rho$ (dotted line).
Lower plot: isotropic pressure $p$ of the polytropic fluid (dashed line),
radial pressure $\tilde p$ (solid black line),
and transverse pressure $\tilde q$ (solid grey line) of the $f(T)$ fluid.}
\end{figure}

\begin{figure}
\begin{center}
\includegraphics{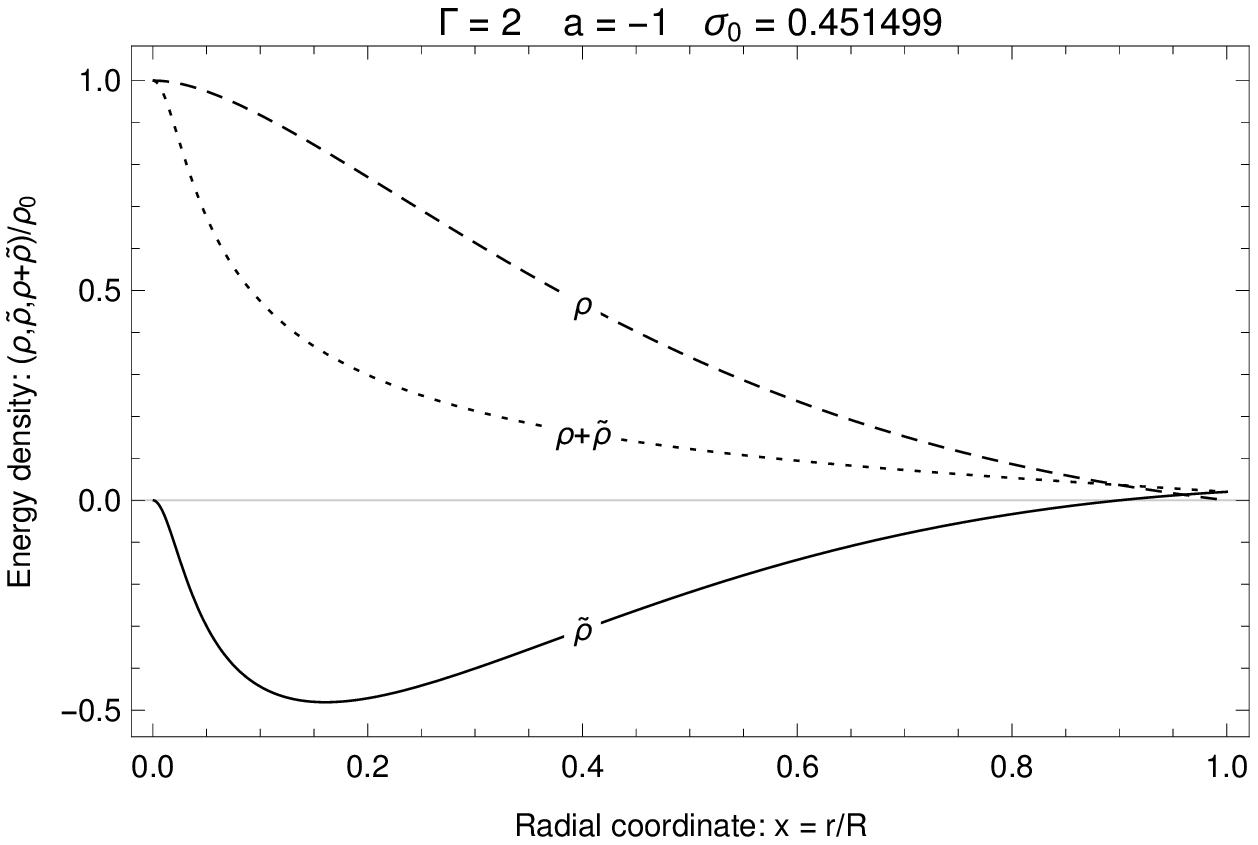}
\caption{\label{fig:minus}
The $\Gamma = 2$ polytrope 
with maximal $\sigma_0$ in $f(T) = T + \frac{\alpha}{2} T^2$-gravity theory,
$a=\alpha\kappa^2=-1$.
Energy density $\rho$ of the polytropic fluid (dashed line),
energy density $\tilde\rho$ of the $f(T)$ fluid in the GR picture (solid line),
and effective energy density $\rho+\tilde\rho$ (dotted line).}
\end{center}
\end{figure}

We begin by examining the energy density and the pressure profiles
in the polytropes obtained with positive values of $a$.
In all configurations, we find qualitative behavior
of the polytropic fluid similar to that in GR;
the fluid has outwardly decreasing energy density,
and at some finite value of the radial coordinate $r=R$
corresponding to the stellar surface,
it satisfies the boundary condition (\ref{eq:bc1}).
This means that the energy density $\rho$
and the pressure $p$ of the polytropic fluid
both vanish at the stellar surface.
Using the GR picture allows us to examine also
the energy density $\tilde \rho$, the radial pressure $\tilde p$,
and the transverse pressure $\tilde q$ of the $f(T)$ fluid.
All these quantities vanish at $r=0$
and are outwardly increasing (positive)
in the central region of the polytrope.
They reach their maxima in the interior of the polytrope
after which they start to decrease, change sign,
and remain negative until reaching the stellar surface.
It is important to emphasize that at the stellar surface
the quantities $\tilde \rho$, $\tilde p$, and $\tilde q$
assume finite negative values
appropriate for joining the vacuum of the $f(T)$-gravity theory,
which differs from the Schwarzschild vacuum of GR.

The energy density and the pressure profiles
of the polytrope obtained with $\Gamma=2$, $a=1$,
and with the maximal value of the central pressure--to--energy density ratio
$\sigma_0 \simeq 0.118$ that could be reached with our numerical procedure
are shown in Fig.~\ref{fig:plus}.
One can observe a very abrupt transition
of the $f(T)$ fluid energy density $\tilde \rho$
and the transverse pressure $\tilde q$
from their ``positive phase'' in the central region of the star
to their ``negative phase'' in the outer region of the star,
while the ``phase transition''
of the radial pressure $\tilde p$ is somewhat smoother.
Comparing the configurations obtained with different values $a$
and maximal $\sigma_0$ that could be reached with our numerical procedures
(corresponding to the upper ends of $N$ vs~$R$ curves in Fig.~\ref{fig:g2_NR}
and with parameters given in Table~\ref{tbl:g2_NR})
reveals that the phase transition becomes sharper as $a$ increases.
However, as our results are obtained numerically,
we cannot claim that a further increase of $\sigma_0$ is in principle possible
and that it would lead to even more steplike phase transitions;
nor can we claim with certainty that such solutions do not exist.
Still, the results that we obtained with positive $a$ and high $\sigma_0$
allow us to loosely divide the interior of such polytropes
into three regions: the core, the phase transition layer, and the halo.
The core of the polytrope is its central region in which,
within the GR picture,
the energy density and pressures of the $f(T)$ fluid are positive
and the effective energy density is approximately constant.
This behavior of effective energy density $\rho+\tilde\rho$
mimics what could be understood as incompressible matter,
i.e., matter that does not allow a further increase of the energy density.
The phase transition layer
is the region in which the approximately constant effective energy density
drops to a considerably lower positive value
and in which the energy density
and the pressures of the $f(T)$ fluid change sign.
We found that, as $a$ or $\sigma_0$ increases,
the radial extent of the phase transition layer becomes narrower.
The halo can be defined as the outer region of the polytrope
in which the $f(T)$ energy density and the pressures are negative
and which ends at $r=R$ where they assume the values
appropriate for joining with the vacuum.

We now turn to the polytropes obtained with negative $a$.
While the energy density of the polytropic fluid is in these solutions
outwardly decreasing and vanishes at the stellar surface,
as in the solutions constructed within GR
or within $f(T)$ gravity with positive $a$,
the behavior of the $f(T)$ fluid is qualitatively different
from that we obtained with positive $a$.
The $f(T)$ fluid energy density vanishes at $r=0$
and is negative throughout the interior of the polytrope,
except near the surface where it becomes positive
in order to match the vacuum at $r=R$.
Figure~\ref{fig:minus} shows the energy density profiles
in the $\Gamma=2$ polytrope obtained with maximal value of $\sigma_0$
allowed by our numerical procedure for $a=-1$.
Similar behavior is found also for the radial and transverse
pressures of the $f(T)$ fluid.
As can be seen from the figure, with negative $a$,
there is no abrupt phase transition
in the $f(T)$ fluid similar to that which is obtained with positive $a$.


\section{Concluding remarks \label{sec:conclusion}}

In order to investigate the features
of $f(T)$-gravity theory in extreme circumstances
such as those arising within highly compact
static spherically symmetric bodies,
we considered the self-gravitating configurations of the polytropic fluid.
In particular, we allowed for the quadratic term in $f(T)$,
i.e., $f(T) = T + \frac{\alpha}{2} T^2$
with positive and negative coefficient $\alpha$.
For the polytropic fluid, we adopted the adiabatic exponent $\Gamma=2$
as it represents the most stiff perfect fluid
which respects the dominant energy condition
and is causal in all pressure regimes.
The numerically constructed solutions revealed that with positive $\alpha$
less matter in terms of number of particles
can be supported against gravity than in GR,
while negative values of $\alpha$ allow a considerably larger
number of particles (see Fig. \ref{fig:g2_NR} and Table \ref{tbl:g2_NR}).
For the interpretation of the solutions,
we adopted the GR picture in which the terms in the equations of structure
that arise due to the quadratic term in $f(T)$
are interpreted as the energy density
and the radial and transverse pressures of the $f(T)$ fluid.
It was found that with positive $\alpha$
the energy density of the $f(T)$ fluid is positive
in the large part of the interior of the polytrope,
while with negative $\alpha$,
its energy density is mostly negative.
This offers a possible explanation of the fact that more particles
of the polytropic fluid can be supported against gravity with negative $\alpha$,
as in that case the $f(T)$ fluid diminishes the effective energy density
(see Fig.~\ref{fig:minus}).
With positive $\alpha$, it was found that within the star
a phase transition of the $f(T)$ fluid occurs,
leading to approximately constant
or even outwardly increasing effective energy density of the polytrope,
which is followed by an abrupt drop (see Fig.~\ref{fig:plus}).
This drop occurs due to the phase transition occurring within the $f(T)$ fluid
in which its energy density and the radial and the transverse pressure
all change sign.

Regardless of the fact that we have constructed
the particle number--to--stellar radius curves,
which are the analog of the well-known mass-to-radius curves
that are usually considered in GR and of which the
maxima indicate the loss of dynamical stability,
at the present, we cannot make any well-grounded claims
about the stability of the polytropes in $f(T)$.
This is in one part due to the fact
that our procedure is static from the outset
and in one part due to not yet settled interpretation
of the stellar mass in $f(T)$ gravity.
A fully dynamical approach would require time-dependent
equations of motion which should be covariant
in the sense that they do not depend on the particular
choice of the tetrad.
Recent progress in formulation of $f(T)$-gravity theory
in the covariant form such as Refs.~\cite{%
   Krssak:2015oua,
   Golovnev:2017dox,
   Hohmann:2018rwf}
makes us hopeful that in our future work we will be able to derive
linearized time-dependent equations of motion
and treat the stability problem perturbatively.
Another issue which remains unresolved is the interpretation
of the mass of a compact in $f(T)$ gravity.
In GR, one obtains the mass
of the compact spherically symmetric static object
directly from the joining of the interior spacetime
with the exterior Schwarzschild metric.
In $f(T)$ gravity, this is at present not possible
since the vacuum metric is not available in closed form
(only a perturbative expression
in the weak-gravity regime is available \cite{DeBenedictis:2016aze}).
Obtaining the mass from the asymptotic behavior of the metric at $r\to\infty$
is also not a completely safe option because it is not clear
to what extent the energy density of the $f(T)$ fluid
contributes to the asymptotic mass.
We intend to address some of these issues in our future work.

\vskip 1em \noindent {\bf Acknowledgements:}
This work is supported by the VIF program
of the University of Zagreb.
The authors thank Andrew DeBenedictis
for reading an early version of the manuscript.


\providecommand{\href}[2]{#2}\begingroup\raggedright\endgroup

\end{document}